\magnification\magstep1
\baselineskip 5.5truemm plus 0.25truemm

\centerline{\bf Summary; Inflation and Traditions of  Research}
\bigskip
\centerline{P. J. E. Peebles}
\centerline{Institute for Advanced Study, Princeton NJ 08540, and}
\centerline{Joseph Henry Laboratories, Princeton University,
Princeton, NJ 08544}
\bigskip\bigskip

There is a considerable spread of opinions on the status 
of inflationary cosmology as a useful
approximation to what happened in the early universe. To some 
inflation is an almost inevitable consequence of well-established
physical principles; to others it is a working hypothesis that has
not yet been seriously tested. I attribute this 
to two traditions of research in physics and astronomy 
that come together in cosmology. The great advances of basic
physics in the 20$^{\rm th}$ century
have conditioned physicists to look for elegance and simplicity.
Our criteria of elegance and simplicity have been informed by the
experimental evidence, to be sure, but the tradition has been
wonderfully effective and we certainly must pay attention to
where it might lead us in the 21$^{\rm st}$ century.
Astronomers have had to learn to deal with incomplete and
indirect observational constraints on complicated systems. The
Kapteyn universe --- a model of our Milky Way galaxy of stars ---
was a product of a large effort of classifying the
luminosities and modeling the motions of the stars 
from the statistics of star counts and star
proper motions and parallaxes. This culminated in a detailed
model that even offered the possibility 
of a ``determination of the amount of dark matter from its
gravitational effect'' (Kapteyn 1922). But as the model was
being constructed people were discovering that it must be
revised, moving us from near the center to the edge of the Milky 
Way, where the phenomenon of stellar streaming
could be reinterpreted as the circular motion of stars in the 
thin disk of the galaxy and the random motions of older stars
in the halo. The rocky road through a vastly richer
fund of observations has led to a picture for the evolution of the
galaxies and the intergalactic medium during the last factor of
five expansion of the universe. Many of the elements are
established in remarkably close detail, but others are
quite uncertain and subject to ongoing debate, 
elegant examples of which are to be found in these 
Proceedings. I think the experience has led to a characteristic
tendency you might have noticed among astronomers, 
to ask first how a new result might have been
compromised by systematic errors in the measurement, or
its interpretation confused by an inadequate model. 
These traditions from physics and astronomy meet in cosmology, 
with predictable consequences for the debate on the status of 
inflationary cosmology. 

There are astronomers who accept inflation as 
persuasively established, and physicists who consider 
inflation speculative. The latter can cite a 
tradition of complexity in physics, as in the
study of condensed matter. Maybe it is not surprising that
several of the most prominent critics of 
the basis for the standard cosmological model
--- the postulate of near homogeneity and isotropy of the
galaxy and mass distributions on the scale of the Hubble 
length --- are condensed
matter physicists. I don't understand why they are not more
influenced by the observational evidence for large-scale
homogeneity, but I can fully appreciate the underlying
question: could the universe really be this simple?

Einstein arrived at the homogeneity picture by
a philosophical argument, that asymptotically
flat spacetime is unacceptable because it is 
contrary to Mach's principle. He considered
this to be more convincing than the empirical 
evidence the astronomers could have given him about our 
island universe of stars and about
the clustered distribution of the brighter spiral nebulae (a
result, we now know, mainly of the concentration of galaxies in
the de Vaucouleurs Local Supercluster). Einstein avoided all these
misleading observational indications; here is an example where
pure thought led to a prediction that proves to agree with 
demanding empirical tests, in the grand tradition of physics. 

It is easy to think of examples of less successful 
application of pure thought, of course.
Most of us would agree that the Einstein-de~Sitter model is the
only reasonable and elegant choice for the parameters of the 
Friedmann-Lema\^\i tre cosmology, because any other
observationally acceptable parameter choice would imply that
we flourish at a special epoch, as the universe is 
making the transition from expansion dominated by the
mass density to expansion dominated by space curvature or a
cosmological constant (or a term in the stress-energy tensor that
acts like one). Maybe this is telling us that the evidence for low
mass density is wrong or somehow incorrectly interpreted. But 
the more likely lesson is that Nature is more complicated 
than we had thought, and we are going to have to revise our
criteria of elegance, as has happened before. 

Inflationary cosmology offers an elegant way to extend 
the standard model for
cosmic evolution back in time to conditions that
cannot be described by the classical Friedmann-Lema\^\i tre
model. But is the early universe really simple in 
the physicists' sense, simple enough that we can hope
to deduce its main properties from what we know about
fundamental physics? Or might there be important elements of the 
complexity that astronomers are used to dealing with, and that we
could hope to unravel only if we were fortunate enough to hit upon 
adequate guidance from the empirical evidence? 

\topinsert
\centerline{TABLE 1}
\centerline{The Case for Inflationary Cosmology}
\medskip
\hbox to \hsize{\hss 
\vbox{
\hrule height 0.6pt
\vskip 2pt
\hrule height 0.6pt
\halign{\strut
        #\hfill & \hfill \ #\ \hfill & \hfill #\hfill \cr
\noalign{\vskip 3pt}
Evidence  & Nature & Status \cr
\noalign{\vskip 4pt}
\noalign{\hrule height 0.6pt}
\noalign{\vskip 4pt}
Compelling elegance of & \lower1ex\hbox{a romantic } 
	& \lower1ex\hbox{seems well}\cr
\noalign{\vskip -4pt}
\quad inflation, and the absence\hskip -0.2truein &\lower1ex\hbox{test}
	 & \lower1ex\hbox{established}\cr
\noalign{\vskip -4pt}
\quad of a viable alternative \cr
\noalign{\vskip 4pt}
Observational case for & a diagnostic,
	& \lower1ex\hbox{preliminary} \cr
\noalign{\vskip -4pt}
\quad flat space sections & not a test? \cr
\noalign{\vskip 4pt}
Observational case for the & \lower1ex\hbox{a diagnostic}
	& \lower1ex\hbox{preliminary} \cr
\noalign{\vskip -4pt}
\quad adiabatic CDM model\cr
\noalign{\vskip 4pt}
Tensor contribution to & a classical 
	& \lower1ex\hbox{open}\cr
\noalign{\vskip -4pt}
\quad the CBR anisotropy & test\cr
\noalign{\vskip 4pt}
Deduction of the  inflaton & & \lower1ex\hbox{a wonderful}  \cr
\noalign{\vskip -4pt}
\quad and its potential & classical & \lower1ex\hbox{dream} \cr
\noalign{\vskip -4pt}
\quad from fundamental physics \cr
\noalign{\vskip 4pt}
 }
\vskip 3pt
\hrule height 0.6pt
\vskip 3pt
}
\hss}
\endinsert

Table 1 is my summary of the pieces of evidence in hand 
that seem to be relevant to this issue. The
characterizations in the second column follow
the physical chemist Wilhelm 
Ostwald,\footnote{$^1$}{As represented in 
the novel, {\it Night Thoughts of a Classical Physicist}
(McCormmach 1982). The title is not meant to distinguish the
protagonist from a quantum physicist; at the time of the story,
1918, there was not yet a quantum theory. In Ostwald's
classification romantic physicists
loved atoms and their curious properties, while
classical physicists distrusted what they considered 
extravagant departures from conventional physics in the 
search for a quantum theory. Ostwald was 
skeptical of the kinetic theory of 
atoms as a basis for heat and chemistry 
(Jungnickel \&\ McCormmach 1986) until 
the experimental success of Einstein's theory of Brownian 
motion won over him and Ernst Mach 
(Whittaker 1953).} who felt that some physicists
have a romantic temperament, eager to pursue the latest ideas, 
as opposed to the classical types who seek to advance knowledge 
by increments from a well established basis of concepts
and methods. In the table I mean by a classical test one that 
follows the old rule of validation by the successful outcome of 
tests of the predictions of a theory. By a diagnostic I mean 
data that can be fit by adjustment of parameters within a model 
for inflationary cosmology. This would turn into a classical test 
if we had another independent constraint on the parameters. 
A romantic test may point to the truth, but it lacks the beauty
of an experimental check of a prediction. 

The first entry in the table refers to the fact that inflationary
cosmology offers an elegant remedy for very real inadequacies
of the classical Friedmann-Lema\^\i tre model. This means we
should pay careful attention to inflation. But those of us
with classical inclinations give more weight to the successful
outcome of predictions than to the ability to devise a theory to
fit a given set of conditions, in what has been termed postdiction. 
It is impressive that no one has come up with an interesting
alternative to inflation, despite the wide advertisement of the ills
it cures. But one may wonder whether the significance is only
that our imagination is limited.

If inflationary cosmology were falsified by a measurement 
that showed that 
space sections of constant world time have nonzero curvature,
then observational evidence for flat space sections could be
counted as evidence for inflation. If inflationary cosmology 
could be adjusted to fit the measured space curvature, then the 
measurement would be a diagnostic of the details of inflation,
not a test. Bharat Ratra, who refined Richard Gott's %(1982)
picture into a well specified model for 
open inflation, %(Ratra \&\ Peebles 1994), 
argues for the latter. Others argue for the former: 
they would take a demonstration
that space curvature is not negligibly small to signify that we
must abandon inflation as it is now understood and search for a
better idea. Still others argue that 
the situation is not that simple: inflationary cosmology  is
more elegant if space curvature is negligibly small, so 
observational evidence that this is the case would be good news, but 
one can work with open inflation, so the discovery of nonzero space
curvature would not be bad news. It would be good if the inflation
community could reach a consensus on which it is before the
astronomers turn the value of space curvature into a postdiction.
It is too soon to settle bets, but the astronomers
seem to be getting close to a useful measurement of
space curvature, as one sees in these Proceedings. 
 
The third entry in the table refers to the 
striking success of the adiabatic CDM model 
for structure formation in fitting 
a broad range of observations. The original motivation for 
the CDM model was not inflation; the model came out of 
a search for a simple way to account for the 
small anisotropy of the thermal cosmic 
background radiation (the CBR). Inflationary cosmology 
offers an elegant explanation
for the initial conditions postulated for the CDM model. This is
encouraging but not a demonstration that processes directly  
related to inflation did provide the seeds for the CDM model. 
It would be no crisis if the CDM model
were found to be wrong; there are other ideas for structure
formation within inflation, or it could be that inflation is not
directly responsible for structure formation. We would have a
critical test if we had an observationally viable model
for structure formation that is not consistent with inflationary 
cosmology, to compare to models inspired by or compatible with 
inflation. As things stand I think we have to count the
observational evidence on how structure formed as a diagnostic of 
(or constraint on) the parameters of inflation, under the
postulate that inflation offers the right picture for the 
early universe.

If the parameters are favorable inflation predicts an
observable contribution to the anisotropy of the CBR from tensor
fluctuations --- fluctuations in the curvature of spacetime.
A detection of the tensor part and a demonstration of 
consistency with a specific model for inflation, perhaps one 
now under discussion, would make believers of most of
us.\footnote{$^2$}{As the experimental success of the atomic
theory of Brownian motion converted Ostwald and Mach.}

As indicated in the last row of the table, the basis for
fundamental physics may become so secure 
that it unambiguously predicts all relevant 
properties of the inflaton. If these properties proved to be
observationally acceptable it would be a prime classical
triumph. It would require a considerable advance in physics, 
but the advances have been prodigious.

We can look at the situation another way by asking where 
cosmology might have been today if we had not had the concept
of inflation. If we thought we had to live with purely 
baryonic matter we would have been pressed to reconcile 
dynamical mass estimates with 
the baryon density required by the successful model for
homogeneous production of light elements
at high redshift, and we would have been pressed
to find a viable model for structure formation. 
We don't know how pressed; maybe we should have worked harder
to save a pure baryonic universe. Nonbaryonic dark matter
matter --- a family of massive neutrinos --- was
considered before inflation. The concept of cold nonbaryonic
matter grew up with inflation, but it surely would have grown 
as well if separated at birth, as did cosmic strings. 
Without inflation people would not have worked so hard to save 
the Einstein-de~Sitter model, and negative space curvature
might have been considered more favorably, but the observations 
would have driven us to about where we are today. Our ideas of
how structure formed have been influenced by inflation, but
did not depend on it. The big difference would be that,
unless we hit on some alternative to inflation, initial 
conditions, including the Gaussian adiabatic fluctuations 
of the CDM model, would have been invoked {\it ad hoc}.
Inflation offers a satisfying way to fill what otherwise 
would be a large gap in our cosmology. Whether or not this
proves to be the true explanation it
certainly has helped drive the present
high level of interest and research in cosmology. 

Without inflation we may not have thought of searching
for the graviton contribution to the CBR anisotropy, which
could yield a believable positive test in 
the classical tradition. Other tests of inflation may 
show up; people are still exploring the possibilities. 
It also is quite conceivable that Nature will not 
be kind enough to give us classical tests of our ideas of
what happened in the early universe; maybe inflation 
is a precursor of a new tradition of research by pure thought. 
Our rules of evidence in science have evolved since 
Newton claimed not to invoke hypotheses, but this would be 
quite a change. The record of forecasts of the end of science 
as we know it leads me to doubt this one, but that is for the
future. These Proceedings document a wonderfully active and 
productive state of cosmology now, in the traditions of
physics and astronomy in their classical and romantic phases. 

\bigskip
The organizers of this conference, Michael Turner and colleagues,
did the community a great service by creating an
exciting and stimulating gathering. My preparation of
this written contribution was aided by discussions with John
Bahcall, David Hogg, Bharat Ratra, and Paul Steinhardt, and 
the work was supported in part at the Institute for Advanced
Study by the Alfred P. Sloan Foundation.

\def\ref{\hangindent=5ex\hangafter=1}
\parskip=0pt
\parindent=0pt
\vskip 0.2in
\centerline{REFERENCES}
\vskip 0.2in

%\ref Gott, J. R. 1982, {\it Nature}, {\bf 295}, 304.

\ref Jungnickel, C. \&\ McCormmach, R. 1986, {\it Intellectual
Mastery of Nature; The Now Mighty Theoretical 
Physics} (Chicago: University of Chicago Press). 

\ref Kapteyn, J. C. 1922, {\it ApJ}, {\bf 55}, 302.

\ref McCormmach, R. 1982, {\it Night Thoughts of a Classical
Physicist} (Cambridge: Harvard University Press).

%\ref Ratra, B. \&\ Peebles, P. J. E. 1994, {\it ApJ}, {\bf 432}, L5.

\ref Whittaker, E. 1953, {\it A History of the Theories of Aether
and Electricity; the Modern Theories} (London: Nelson and Sons),
reprinted 1960 (New York: Harper Torchbooks).

\bye